# Novel Increase of Superconducting Critical Temperature of an Iron-Superconductor due to Ion Implantation

Kriti R Sahu[1†,2] Thomas Wolf[3] A K Mishra[4] A Banerjee[4] V Ganesan[4,5] Udayan De[2*]

[1]*Physics Dept., Bhatter College, Dantan, Paschim Medinipur, W Bengal, India 721426*
[2]*Physics Dept., Egra S. S. B. College, Egra, Purba Medinipur, W Bengal, India 721429*
[3]*Institute for Quantum Materials and Technologies, Karlsruhe Institute of Technology, Karlsruhe D-76021, Germany (Retired)*
[4]*UGC-DAE Consortium for Scientific Research, Indore, M.P., India 452001*
[5]*Medi-Caps University, A.B. Road, Pigdamber, Rau, Indore, M.P., India453331*

Energetic ion irradiation usually decreases superconducting critical temperature($T_c$), with the few exceptions involving increases up to a few K only. However, our recent $2.5 \times 10^{15}$ Ar/cm$^2$ irradiations by 1.5 MeV Ar$^{6+}$ enhanced $T_c$ of the single crystal Fe-superconductor Ba(Fe$_{0.943}$Co$_{0.057}$)$_2$As$_2$ by 8.2 K from its initial onset $T_c$ of ~16.9 K as measured from the real part of the magnetic susceptibility, matching measurements from the imaginary part, electrical resistivity and magnetization. Ozaki et al. (2016) explained their $T_c$ increase of 0.5 K in FeSe$_{0.5}$Te$_{0.5}$ films with the thickness (t) < the irradiating proton range (R), as due to a nanoscale compressive strain developed from radiation damage of the lattice. Here, Ar irradiation with t > R results in an Ar implanted layer in our crystal. Implanted inert gas atoms often agglomerate into high-pressure bubbles to exert a large compressive strain on the lattice. We suggest that this additional compressive strain could be the reason for such a large (~49%) $T_c$ increase.

## 1. Introduction

Superconductivity originates in BCS-type metallic superconductors in the metallic environment and in oxide HTSCs (High Temperature Superconductors) in the Cu-O layers. Its origin in magnetic layers of Fe-As or Fe-Te/Se in Fe pnictide / chalcogenide superconductors or Fe-HTSCs are novel, involving a few exotic superconducting states.[1-6] More research on the mechanism of superconductivity is still needed. On the application side, these Fe superconductors have a good potential of being fabricated into practical wires or cables for high field magnets due to fabrication advantages, and 58 K or higher $T_c$ in a few

Fe-HTSCs.[2,7] Such magnets may be used in radiation environments of accelerators and fusion reactors. Considering these application-oriented aspects, radiation effects in Fe-HTSCs are being studied fairly widely.[8] To join the radiation damage documentation experiments, we have chosen the well-studied Co-doped iron pnictide of $Ba(Fe_{(1-x)}Co_x)_2As_2$, with x = 0.057, a composition slightly below the optimum doping concentration – but in single crystal form to eliminate grain boundary complications. We decided to work on bulk samples to avoid thin films that often have special properties. Here, irradiation by heavier ions ($Ar^{6+}$), specifically $2.5×10^{15}$ Ar-ions/cm$^2$ irradiation with 1.5 MeV $Ar^{6+}$ ions, instead of lighter ions have been selected, to inject higher damage energy per (lattice) atom (dea) and hence to have more effect on the superconducting properties. Our motivation has been to get larger irradiation-induced changes of $T_c$ that get more attention, but certainly without anticipating any increase of $T_c$. However, the choice of the ion beam made the ion range (R) smaller than the sample thickness (t = ~ 200 μm).

Quite surprisingly, the above-mentioned Ar ion implantation caused $T_c$ to increase and that too by a whopping ~ 49% or ~ 8 K, confirmed from measurements of electrical resistivity, magnetization and the real and imaginary components of magnetic susceptibility. This is in stark contrast to 52 out of 55 references listed in Table 1 of the review of M. Eisterer reporting damage or decrease of $T_c$ of different Fe-based superconductors.[8] The reported increase at best has been 0.5 K in the review.

In terms of timeline, an invited talk[3] in Berlin (2019) followed by an early 2020 talk[4] in Kolkata have been our preliminary reporting, basically to invite discussions. The present work additionally explores a possible cause or causes of such exceptionally large increase of $T_c$, after summarizing present radiation modifications of electrical resistance, magnetic susceptibility ($\chi'$ and $\chi''$) and magnetization (M). As stated in the Abstract, Ozaki et al. showed that nanoscale lattice strain induced by radiation damage in their t < R samples is increasing $T_c$ in these compressed nano-regions with the proximity effect turning it into a bulk effect.[9] Here, we point out tuning of $T_c$ in $Ba(Fe_{(1-x)}Co_x)_2As_2$ by bulk pressure.[10]

Inert gas atoms are insoluble in solids. So, on implantation into a solid, these often precipitate forming small-sized high pressure (HP) bubbles in the solid.[11-13] This topic has been widely studied for decades, particularly for the inner walls of nuclear reactors. The state (solid, liquid or gaseous) of the HP bubbles in the solid can be anticipated from the Ar pressure-temperature phase diagram, cited here.[13] Faraci et al. (1991) and Felde et al. (1984) could demonstrate the presence of HP solid Ar bubbles.[11,12] The present work proposes, for the first time, that such HP bubbles, possible in our t > R case of Ar implantation, exert a large additional pressure on the Ba(Fe$_{(1-x)}$Co$_x$)$_2$As$_2$ lattice, concomitantly causing a large additional increase of T$_c$.

## 2. Material and Methods

Single crystals (SXLs) of Ba(Fe$_{(1-x)}$Co$_x$)$_2$As$_2$, x = 0.057, were grown by the self-flux method.[2,5] XRD of UR (Un-Radiated) samples in Fig.1 shows only (00l) reflections from the large face of our 1.7 – 3 mm (length), 1 – 2 mm (width) and 0.2 – 1 mm (thickness) samples and no x-ray detectable second phase. This proves the purity and single crystalline nature of the samples.

Here, 200 μm thick SXLs have been irradiated at the Inter-University Accelerator Centre (IUAC), New Delhi, India, by a 1.5 MeV Ar$^{6+}$-beam. The crystals have been characterized first in UR and then in AR (After Radiation) condition. All characterizations have been summarized in Table I.

A Quantum Design SQUID magnetometer has been used for the measurement of the real and imaginary parts of the magnetic susceptibility (Fig. 2).

Electrical resistance for (a,b) plane conduction has been measured (Figs. 3 and 4) using a standard DC 4-probe technique under H = 0 & 4 Tesla magnetic fields in a 14 T PPMS system from Quantum Design, USA, H being applied along the c axis direction of the sample. For this, rectangular-shaped samples have been cut with a wire-saw. Electrical contacts for conduction along the ab-plane have been made by attaching thin copper wires with silver epoxy. The larger uncertainty in resistivity value than in normalized or relative resistivity value is likely due to the larger uncertainty in the measurement of sample dimensions (compared to electrical measurements). So, Figs.3 and 4 present normalized resistivity vs. temperature (T) data, as is often done in the literature.[9]

Figs. 5 and 6 show ZFC (Zero Field Cooled) and FC (Field Cooled) magnetization (M) vs. temperature (T) in the crystals before and after the irradiation as measured in a Quantum Design SQUID magnetometer.

## 3. Theoretical Considerations

Radiation damage test data ideally include, in addition to fluence, $\Phi t$, either displacements per atom (dpa) or damage energy per atom (dea or DEPA). For comparing different types of irradiations, the mean energy, $E_{p,av}$, transferred by any irradiation to an atom of the target lattice, has often been used as a measure of the effective irradiation instead of the fluence.[14,15] This is the damage energy per atom (dea). This parameter has been successfully utilized in the radiation damage investigation of different materials including superconductors.[8,14,15] Before elaborating on the increase in $T_c$ via inert gas bubble formation, details of the higher effectiveness (in terms of $E_{p,av}$) of the present Ar ion irradiation with respect to the 190 keV proton irradiation will be presented.[8] Defects can be generated by elastic collisions, only if the transferred energy exceeds the binding energy, $E_d$, of the lattice atom (typically a few eV in metals and 10 – 40 eV in ionic crystals).[8] Maximum Energy transfer in a head-on elastic collision is: $T_m = T_{(max)} = 4M_l \cdot M_p \times E_p / (M_l + M_p)^2$ for the projectile (ion / neutron) with mass ($M_p$) and energy ($E_p$) falling on a lattice with atoms of mass $M_l$.

For charged particle irradiation with low energy transfer, $E_{p,av}$ can be calculated considering only energy transfers higher than the threshold energy $E_d$ as follows.[14,15]

$$E_{p,av} = \frac{T_m E_d}{(T_m - E_d)} ln\left(\frac{T_m}{E_d}\right) \pi Z_l^2 Z_p^2 e^4 \left[\frac{M_l M_p}{(M_l + M_p)^2}\right] \frac{(T_m - E_d)}{T_m E_d} \frac{\Phi t}{E_p}$$

$$= ln\left(\frac{T_m}{E_d}\right) \pi Z_l^2 Z_p^2 e^4 \left[\frac{M_l M_p}{(M_l + M_p)^2}\right] \frac{\Phi t}{E_p} \quad \ldots \quad \ldots \quad \ldots \quad (1)$$

## 4. Results

*4.1 Results and Discussion*

XRD showed a gradual decrease of the lattice parameter c due to increasing Co-substitution: c = 13.031 Å for x = 0.000, c = 12.999 Å for x = 0.057 & 12.988 Å for x = 0.102. A broadening of XRD peaks indicates the generation of lattice defects. Large broadening for irradiation by the high fluence of $10 \times 10^{15}$ Ar-ions/cm$^2$ in Fig. 1 shows significant lattice defects. XRD after $2.5 \times 10^{15}$ Ar-ions/cm$^2$ irradiation shows no such drastic damage of the structure. Results for the magnetic susceptibility (Fig. 2), relative electrical resistivity (Figs. 3 and 4) and ZFC magnetization (Figs. 5 and 6) for the present sample before and after our irradiation, and similar literature data, along with some energy transfer calculation (see Section 4.3.2) results, have been given in Table I, to help the discussion. As expected, neither the real part ($\chi'$), nor the imaginary part ($\chi''$) of magnetic susceptibility showed (Fig. 2) superconductivity in the x = 0 sample.

In Fig. 2 and Table I, for the 5.7% Co-doped Ba(Fe,Co)$_2$As$_2$ sample (called 2Ba5.7Co in UR and AR conditions), our irradiation increases T$_c$($\chi'$), as given by the $\chi'$ step, from 16.9 K to 25.1 K. This implies an increase by 8.2 K in T$_c$($\chi'$) due to our Ar-irradiation, a case of ion implantation. From SRIM (Stopping and Range of Ions in Matter) calculation, we estimate a range (R) of 0.903 μm for a 1.5 MeV Ar beam in our sample. So, a buried Ar-implanted layer is expected to be created in our sample. The inset of Fig. 2 shows only one peak in $\chi''$ vs. T graph as it should be for a single crystal, while pellet or granular samples usually show inter-grain and intra-granular peaks. Here, our irradiation increases T$_c$($\chi''$) from 16.0 K to 24.0 K, implying an increase of 8 K. There is thus, an agreement between the measurements on the imaginary part ($\chi''$) and the real part ($\chi'$) of the magnetic susceptibility.

In addition, the independent measurement of T$_c$ from electrical resistivity also records that the 2.5×10$^{15}$ Ar-ions/cm$^2$ irradiation increases (Figs. 3 and 4) T$_c$(completion) from 16.0 K to 23.8 K i.e., by 7.8 K. This is in good agreement with the magnetic results.

The application of a magnetic field is seen to shift the superconducting transition for unirradiated as well as irradiated samples significantly towards lower temperatures, as expected in superconductors (Figs. 3 and 4).

It is satisfying that the ZFC magnetization data in Figs. 5 and 6 indicate an increase of T$_c$(onset), matching the resistivity and magnetic susceptibility results.

*4.2 Existing Explanations of T$_c$ Increase by Irradiations*

On searching the literature for explanations of the irradiation-induced T$_c$ increase cases observed in different superconductors, 4 explanations have been found.

Explanation 1: Irradiation-generated nanoscale compressive strain in the lattice caused localized T$_c$ increase at different spots, and their linkage by the proximity effect resulted in 2.8% increase of T$_c$ in t < R Fe-superconductor of Ozaki et al., as already discussed.[9]

Explanation 2: Marginal rise of T$_c$ in low-T$_c$ A-15 superconductors (Mo$_3$Ge and Mo$_3$Si) due to the irradiation-induced smearing or broadening of the DOS (Density of States) peak in the DOS = N(E) vs. E plot is known.[16]

Explanation 3: For Ba$_2$Sr$_2$CaCu$_2$O$_{8+x}$ (Bi-2212), T$_c$ vs. O-content graph is an inverted parabola, and the major effect of radiations like our 55 MeV Li$^{3+}$ irradiation was removal of O-atoms.[17] So, the same 55 MeV Li$^{3+}$ irradiation increased T$_c$ (by 4.0 K) in samples with higher than optimum O-content, and decreased T$_c$ (by 11.5 K) in samples with optimized O-content.[17]

Explanation 4: Teknowijoyo et al. proposed that their observed increase of $T_c$ (by ~ 4.54%) in FeSe due to 2.5 MeV electron irradiation was due to local strengthening of the pair interaction by irradiation-induced Frenkel defects.[8,18]

One or more of these and/or a new Explanation will be applicable to the present observation of $T_c$ increase. Often irradiation results have been complex and apparently conflicting. Mizukami et al. found that on 2.5 MeV electron irradiation of $BaFe_2(As_{1-x}P_x)_2$, $T_c$ increases while $T_N$ (temperature of antiferromagnetic ordering) decreases for underdoped samples (x = 0.16 and 0.24).[19] Here, $T_c$ was enhanced (by ~ 2 K or so only, as shown in their Fig. 3(b)), only up to the irradiation dose of ~ 3 $C/cm^2$. But $T_c$ is depressed at high dopant concentrations (x ≳ 0.28). However, the Eisterer group observed $T_c$ to decrease by irradiation in underdoped $BaFe_2(As_{0.76}P_{0.24})_2$ crystals. Unseen factors may have caused this difference.[8]

*4.3 Tentative Explanation of the Larger Improvement of $T_c$ by the Present Irradiation*
*4.3.1 General*

None of the 4 explanations can explain the above-discussed ~8 K increase of $T_c$ in our SXLs of $Ba(Fe_{0.943}Co_{0.057})_2As_2$ due to $2.5 \times 10^{15}$ $Ar/cm^2$ irradiation by 1.5 MeV $Ar^{6+}$. Compression of the lattice by radiation defects increased $T_c$ by 2.8% in Explanation 1. But the unique t > R condition here allows ion implantation and possible agglomeration of Ar to form high pressure solid Ar precipitates (as bubbles).[11-13] This can exert, as discussed below, additional compression of much higher magnitude to account for the much larger $T_c$ increase.

Table I shows that the sample thickness (t) is larger than the range (R) of the irradiating particle only in our experiment. This is a necessary condition for the formation of an implantation layer, at a depth R. Ions will fully penetrate the sample and go out in cases where t < R (in Table I), resulting in no possibility of implantation and bubble formation. These cases may show the smaller $T_c$ increase only due to lattice-compression resulting directly from the radiation defects. Such lattice-compression has, in fact, been observed by HRTEM.[9] Our experiment might have produced (i) lattice-compression by the passage of the particle irradiation as in Ozaki et al. work, and (ii) lattice-compression due to the pressure of high-pressure inert gas bubbles in the implantation layer, with (ii) as the main contribution.[9]

As a support to this conjecture of implanted Ar-ions agglomerating into high-pressure Ar bubbles in our samples, let us recall such experiments finding HP bubbles in similar lattice systems.[11-13] This process should be able to exert substantial pressure also in case of the Ba(Fe,Co)$_2$As$_2$ lattice. It is well known in the literature and already discussed in the Introduction section that high pressure, like chemical doping, can generate superconductivity or enhance T$_c$.[10] For example, high pressure in a special diamond anvil cell generated superconductivity in SrFe$_2$As$_2$ (giving maximum T$_c$ of 27 K at around 4 GPa) and also in BaFe$_2$As$_2$ (giving maximum T$_c$ of 29 K at around 4 GPa).[23] The Karlsruhe group showed, in their Figs. 2 (a) and (b), 15 K increase of T$_c$(p) of underdoped (x = 0.041) Ba(Fe$_{(1-x)}$Co$_x$)$_2$As$_2$ under an applied hydrostatic pressure, p, of ~ 3.5 Gpa.[10] These are recalled to stress that pressure effect is capable of causing the large T$_c$ rise that we observe.

Ozaki et al. referred to an earlier TEM observation of strain fields created in Y-123 single crystals by low energy proton and neutron irradiations.[9,24,25] Ozaki et al. further observed, in their own HRTEM and STEM images, inwardly curved lattice fringes, which indicated the presence of a strain field around the cascade defects in their FeSe$_{0.5}$Te$_{0.5}$ films (with unirradiated T$_c$ of 18 K).These defects originated from their 190 keV proton irradiation.[9] Internal compressive strain, especially along the (a,b)-plane directions, has been experimentally found to enhance T$_c$ in FST films, as detailed in Refs.[16,41-43] of Ref.8.[8] Such situations have been theoretically studied by Bellingeri et al.[26,27] For a superconducting film on a substrate, substrate-induced strain affects T$_c$, providing another example of the pressure effect.[27]

*4.3.2 Damage Energy per Atom ($E_{p,av}$) Comparison*

(1) Calculation of $E_{p,av}$ for Ba(Fe$_{0.943}$Co$_{0.057}$)$_2$As$_2$, irradiated by Ar-beam:

$E_{p,av}$ = 98.74 eV and $T_m$ = 1.334 MeV have been calculated, taking the values of $Z_p$(Ar) = 18, $Z_l$ = [(56×1 + 26×1.886+27×0.114+33×2)/5] = 34.8228, $M_p$(Ar) = 39.962, $M_l$ = [(137.327×1 + 55.847×1.886+58.933×0.114+74.992×2)/5] = 79.8714 and $E_d$ = 40 eV. Φt = 2.5×10$^{15}$ cm$^{-2}$, $E_p$ = 1.5 MeV

(2) Calculation of $E_{p,av}$ for FeSe$_{0.5}$Te$_{0.5}$, irradiated by proton beam:[9]

$E_{p,av}$ = 0.0275 eV and $T_m$ = 9.389 keV have been calculated, taking the values of $Z_p$(proton) = 1, $Z_l$ = [(26×1 + 34×0.5+52×0.5)/2] = 34.5, $M_p$(proton) = 1.008, $M_l$ = [(55.847×1 + 78.96×0.5 + 127.6 × 0.5)/2] = 79.5635 and $E_d$ = 40 eV. Φt = 1×10$^{15}$ cm$^{-2}$, $E_p$ = 190 keV

(3) Calculation of $E_{p,av}$ for FeSe, irradiated by electron beam:[18]

$E_{p,av}$ = 15.32 eV and $T_m$ = 81.39 eV have been calculated, taking the values of $Z_p$(electron) = 1, $Z_l$ = [(26 + 34)/2] = 30, $M_p$(electron) = 5.486×10$^{-4}$ amu, $M_l$ = [(55.847×1 + 78.96×1)/2] = 67.404 and $E_d$ = 40 eV. Φt = 1.12×10$^{19}$ cm$^{-2}$, $E_p$ = 2.5 MeV electron.

Table I includes the above-calculated $T_{(max)}$ and $E_{p,av}$ for earlier and our irradiations. Here, we find that our radiation on Ba(Fe$_{0.943}$Co$_{0.057}$)$_2$As$_2$ has dea or $E_{p,av}$ (mean energy transferred by the irradiation to an atom of the target lattice) higher than that of the three other charged particle irradiations.[9,18] Based on the concept of lattice compression due to radiation damage only, this higher dea can give somewhat higher $T_c$ rise, but not the large ~ 8 K rise.

*4.3.3 Discussion of $T_c$ Increase by High-Pressure Bubbles*

Being insoluble, inert gas atoms coming inside a solid by ion implantation or through nuclear reaction in a reactor often tend to agglomerate and form HP bubbles, as already mentioned. All measurements in the literature show the pressure built up in the bubbles to be very high, although quantitative results of different investigations differ. According to the phase diagram, Ar gas condenses forming a solid phase at pressures above 1 GPa, even at a temperature as high as 300 K.[13] The following data on bubbles are room temperature values. These will somewhat change on cooling due to difference in contraction of the lattice and condensed bubbles. In our Ref.11, Felde et al. (1984) measured pressure, P, for Ar implantation in Al: P(from energy shift of valence excitations) = 6(3.5) GPa and P (from TEM diffraction patterns) = 3 GPa, in r =15 Å bubbles. Faraci et al. (Ref. 12) found evidence for crystalline over-pressurized Ar clusters in Al (2.5 GPa) and Si (4.4 GPa). Here, we add that Fe-superconductors are metallic in nature. Faraci et al. extended their work in 1997 to high-pressure Kr clusters in Be and Si using XANES. Smaller ions like H and He behave somewhat differently on implantation. Still, we note that Djukica et al. (2007) probed elastic strains in the compound LiNbO$_3$ due to He-implantation. We, therefore, understand that inert gas ion implantation can form HP bubbles in metals and semiconductors to compress the lattice around the bubbles. Same should be possible for Ar implantation in Ba(Fe,Co)$_2$As$_2$. Such compressed regions should have enhanced $T_c$.

Reports like the Karlsruhe report of ~15 K increase of $T_c$ under a pressure of ~ 3.5 Gpa in underdoped Ba(Fe$_{(1-x)}$Co$_x$)$_2$As$_2$ and the above-expected high pressure of Ar bubbles formed in present samples by Ar-implantation, should lead to large enhancement of $T_c$ in isolated compressed regions around the HP bubbles in present samples.[10] We will go to the details after outlining the explanation of Ozaki et al.[9]

Fig.3 of Ref.9 is the experimentally obtained internal strain map of p-irradiated FeSe$_{0.5}$Te$_{0.5}$ films (with unirradiated $T_c$ of 18 K) of Ozaki et al., and their perspective view of the resulting spatial distribution of enhanced $T_c$.[9] They experimentally found their irradiated film to consist of two types of "nano-regions": (highly compressed) high-$T_c$ regions (with $T_c$ reaching 25 K) and (highly stretched) low-$T_c$ regions (with $T_c$ dropping up to 15 K), the regions being linked in a cobweb-like network.[9] The greater volume fraction of the compressive strain regions and proximity effect resulted in the enhanced bulk $T_c$ of 18.5 K. Our observation of ~ 8 K increase of bulk $T_c$, in contrast to the above-discussed ~ 0.5 K increase, is interestingly larger. This large increase can now be attributed to implanted Ar agglomerating into high-pressure precipitates in form of bubbles in the implantation region of our sample. Superconducting critical temperature of our Ba(Fe, Co)$_2$As$_2$ sample was $T_c(\chi'') =$ 16 K before the Ar-implantation. Localized highly compressed regions around these high pressure Ar bubbles in the irradiated samples will have much enhanced superconducting critical temperature, $T_c(\chi'') \gg$ 16 K, in a lower $T_c$ surrounding. This is similar to above-mentioned perspective view of Ozaki et al., but expected to involve much higher pressure as already discussed. Such enhanced $T_c$ regions can link up by proximity effect if their separations are shorter than the coherence length, to give the observed bulk $T_c(\chi'')$ of 24 K. Future characterizations similar to that of Ozaki et al. can experimentally confirm the presently offered explanations, while the irradiation-enhancement of bulk $T_c$ has been presently confirmed by measuring four superconducting parameters.

## 5. Conclusion

Superconducting Critical Temperature ($T_c$) was measured for ~ 200 μm thick single-crystal samples of a popular iron pnictide superconductor, Ba(Fe$_x$Co$_{(1-x)}$)$_2$As$_2$, for x = 0.057 (underdoped), before and after 1.5 MeV Ar$^{6+}$ irradiation to a fluence of 2.5 × 10$^{15}$ Ar/cm$^2$. Measurements of magnetic susceptibility ($\chi'$ and $\chi''$), electrical resistivity ($\rho$), and magnetization convincingly show that this irradiation surprisingly increased $T_c$ by ~ 49% or ~ 8 K. This is in stark contrast to the usual decrease of $T_c$ by different irradiations in different superconductors. Also, $T_c$ increase, if any, has been much smaller in the few earlier observations of $T_c$ increase in Fe-HTSCs.

In fact, ion or other radiation usually damages or degrades superconductivity sustaining features like the structure of a superconducting material, leading to a decrease of $T_c$. So, an overall increase of $T_c$ by irradiation has to involve new phenomena. One new phenomenon was lattice compression in nano-regions due to proton radiation damage in t < R films of FeSe$_{0.5}$Te$_{0.5}$ with $T_c$ = 18K. However, this increased the bulk $T_c$ from 18 K to 18.5 K only.[9] Our ~8 K increase of $T_c$ can be explained by invoking an additional source of high internal pressure.

Among the reported irradiations, our experiment alone allows ion implantation. It is likely,

as already discussed, that implanted Ar forms high-pressure miniature bubbles that produce additional and large compression of the lattice. In fact, internal pressure up to 6 GPa has been reported for Ar bubbles in metals.[20] We have also pointed out that $T_c$ of an x = 0.041(underdoped) crystal increased from 10 K to 25 K under a pressure of 3.5 GPa.[10] Therefore, we propose that the present Ar-implantation with self-agglomeration of Ar into high-pressure solid Ar bubbles, has formed nano-regions of highly enhanced $T_c$. These high $T_c$ nano-regions link up through the proximity effect and lead to a fairly high enhancement (~ 8 K) of the bulk $T_c$. To our knowledge, neither such large radiation-induced increase of bulk $T_c$ in Fe HTSCs nor the explanation based on high-pressure bubble formation of implanted inert atoms has been discussed by other authors in the literature. Obviously, more ion implantation (t > R) and radiation damage (t < R) experiments with microstructural and superconducting characterizations of pre- and post-irradiated Fe-HTSC samples are necessary for a more confirmed understanding. Such large $T_c$ enhancement may be discovered in other superconductors as well. On the application side, inert gas ion implantation with HP bubble formation may lead to the advantages of higher superconducting transition temperatures.

**Acknowledgment**

Cooperation of the IUAC authority and researchers in ion irradiation and XRD characterizations is gratefully acknowledged. We thank UGC-DAE CSR, Indore, India, for funding the research stay for one of us (KRS) to take a few measurements. Partial support by the Humboldt Foundation, Bonn, Germany, and KIT, Karlsruhe, towards a short discussion stay of UD at KIT is gratefully acknowledged. It is a pleasure to acknowledge discussions with Prof. Sandip Sengupta (UML, USA) and Profs. P. K. Giri (IIT, Guwahati), P. Roy (Inst. of Child Health), M. Nanda Goswami (Midnapore College (Autonomous)) and D. Mohanta (Tezpur Univ.) of India.

*Email: udekol61@gmail.com
†Email: kriti.basis2020@gmail.com

**Figure captions:**

Fig. 1(a) X-Ray Diffraction (XRD) pattern of plate-like single crystal $Ba(Fe_{1-x}Co_x)_2As_2$, $x = 0.057$ samples – in UR (Un-radiated) & AR (After Radiations, by 1.5 MeV Ar-beam) conditions. The samples show only (00l) reflections. Peaks in the sample "7Ba 57 AR" with high dose ($10\times10^{15}$ ions/cm$^2$) irradiation are highly broadened with respect to the peaks in un-irradiated sample (called "Ba 57 UR") and the lower dose ($2.5\times10^{15}$ ions/cm$^2$) irradiated sample (called "5Ba 57 AR")

Fig. 1(b) XRD in 2θ range 56° to 57.5° shows the (008) peak in details.

Fig. 2 Real part of magnetic susceptibility ($\chi'$) (i) for the un-radiated pure $BaFe_2As_2$ sample (called Ba122 UR sample, indicated by black solid sphere); (ii) un-radiated 5.7% Co-doped $BaFe_2As_2$ sample (called Ba5.7Co UR sample), shown by red-colored half-filled rectangle; and (iii) 5.7% Co-doped sample after radiation to $2.5 \times 10^{15}$ Ar-ions.cm$^{-2}$ (called "Ba5.7CoAR(2.5)" sample), indicated by blue-colored half-filled diamond symbols. Inset shows imaginary ($\chi''$) part of magnetic susceptibility, with the un-radiated sample indicated by brown circles (The inset is taken from our conference presentation).[4]

Fig. 3 Temperature-dependent relative electrical resistivity along (a,b) plane at different magnetic fields (0 T and 4 T), applied along the c axis direction, for single crystals of un-radiated $Ba(Fe_{0.943}Co_{0.057})_2As_2$ (called "Ba5.7Co UR" sample).

Fig. 4 Temperature-dependent relative resistivity along (a,b) plane at different magnetic fields (0 T and 4 T), applied along the c axis direction, for single crystals of After-Radiated $Ba(Fe_{0.943}Co_{0.057})_2As_2$ (called "Ba5.7Co AR(2.5)" samples). Fluence of this 1.5 MeV $Ar^{6+}$ irradiation has been $2.5 \times 10^{15}$ cm$^{-2}$.

Fig. 5 (a) Temperature-dependent magnetization of the "Ba5.7Co UR" SXL sample (un-radiated $Ba(Fe_{0.943}Co_{0.057})_2As_2$), measured in magnetic fields of 0 Oe (ZFC- Ba5.7Co UR) and 100 Oe (FC-Ba5.7Co UR – 100Oe). (b) An enlarged view of the superconducting transition region.

Fig. 6(a) Temperature dependent magnetization (M) of the "Ba5.7Co AR (2.5)" SXLsample

14$(Ba(Fe_{0.943}Co_{0.057})_2As_2)$, after radiation (AR) with a dose $2.5 \times 10^{15}$ cm$^{-2}$), measured in magnetic fields of 0 Oe (ZFC- Ba5.7Co AR(2.5)) and 100 Oe (FC-Ba5.7Co AR(2.5) – 100Oe). (b) An enlarged view of the superconducting transition region.

**Table I**. Consolidated results on $T_c$ (superconducting transition temperature) increase due to present and earlier particle irradiations of Fe-HTSCs. UR (Un-radiated) and AR (After Radiation) $T_c$ are listed, along with the Dose and Projected Range (R) of the fast particle beam, Maximum Energy Transfer ($T_m$) in a head-on elastic collision, and Damage Energy per Atom (dea = $E_{p,av}$), with calculation details given in the text.

| Sample | Sample thickness (t), note | Particle | Dose (cm$^{-2}$) | $T_m$ (keV) | dea = $E_{p,av}$ (eV) | $T_c$(UR) in K | $T_c$(AR) in K | Projected range (R) in sample | Ref. |
|---|---|---|---|---|---|---|---|---|---|
| FeSe$_{0.5}$Te$_{0.5}$ | R >> t | n = fast neutrons | 1.8 × 10$^{17}$ | -- | -- | 14.4 | 14.4 | -- | [8] |
| FeSe$_{0.3}$Te$_{0.7}$ | R >> t | n = fast neutrons | 1.8 × 10$^{17}$ | -- | -- | 14.2 | 14.35 | -- | [8] |
| FeSe$_{0.5}$Te$_{0.5}$ | 100 nm R > t | 190 keV protons | 1 × 10$^{15}$ | 9.389 | 0.028 | 18.0 | 18.5 | 1.12 μm | [9] |
| FeSe | 0.03 mm R > t | 2.5 MeV electrons | 1.12 × 10$^{19}$ | 0.081 | 15.323 | 8.8 | 9.2 | 1.22 mm | [15] |
| Ba(Fe$_{0.943}$Co$_{0.057}$)$_2$As$_2$ #1, χ′ data | ~ 200 μm, t > R | 1.5 MeV Ar$^{6+}$ | 2.5× 10$^{15}$ | 1.334k | 98.74 | 16.9 | 25.1 | 0.903 μm | Our work |
| Ba(Fe$_{0.943}$Co$_{0.057}$)$_2$As$_2$ #2, χ″ data | ~ 200 μm, t > R | 1.5 MeV Ar$^{6+}$ | 2.5× 10$^{15}$ | 1.334k | 98.74 | 16.0 | 24.0 | 0.903 μm | Our work |
| Ba(Fe$_{0.943}$Co$_{0.057}$)$_2$As$_2$ #3, ZFC M data | ~ 200 μm, t > R | 1.5 MeV Ar$^{6+}$ | 2.5× 10$^{15}$ | 1.334k | 98.74 | 17.0 | 25.0 | 0.903 μm | Our work |
| Ba(Fe$_{0.943}$Co$_{0.057}$)$_2$As$_2$ #4, ρ data, at 0 T | ~ 200 μm, t > R | 1.5 MeV Ar$^{6+}$ | 2.5× 10$^{15}$ | 1.334k | 98.74 | 16.0 | 23.8 | 0.903 μm | Our work |
| Ba(Fe$_{0.943}$Co$_{0.057}$)$_2$As$_2$ #5, ρ data, at 4 T | ~ 200 μm, t > R | 1.5 MeV Ar$^{6+}$ | 2.5× 10$^{15}$ | 1.334k | 98.74 | 10.0 | 21.2 | 0.903 μm | Our work |

#1 = $T_c$(onset) from χ′ measurement, #2 = $T_c$(peak) from χ″ measurement, #3 = $T_c$(onset) from ZFC magnetization (M) measurement, #4 and #5 = $T_c$(completion) from resistivity(ρ) measurement at 0 and 4 Tesla, respectively.

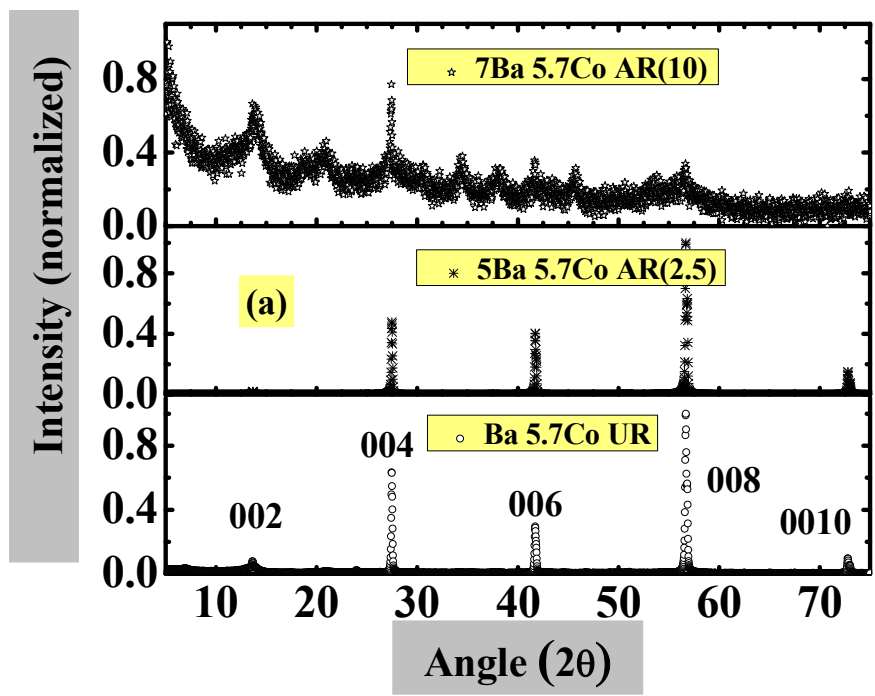

Fig. 1(a)

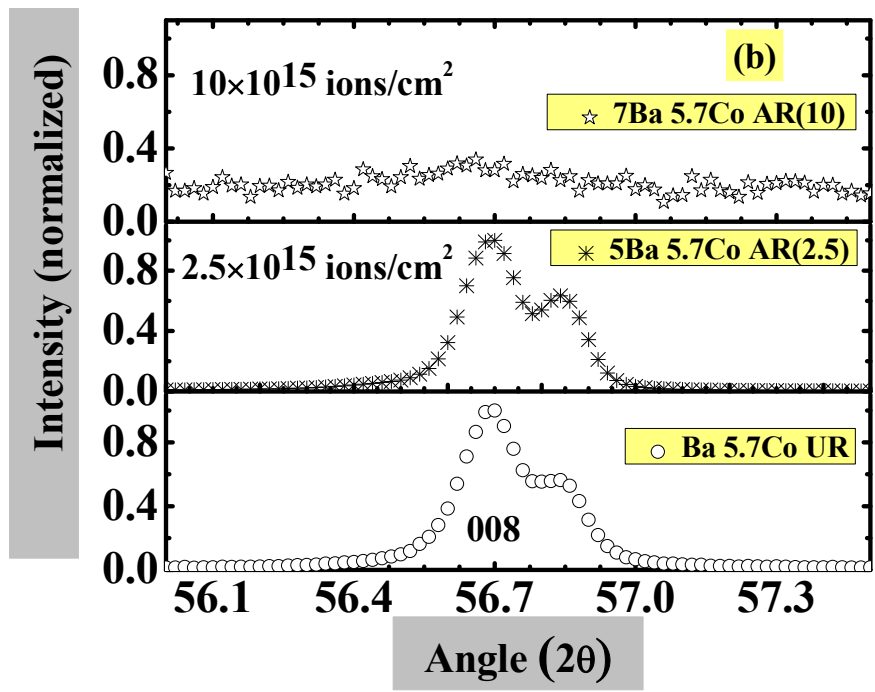

Fig. 1(b)

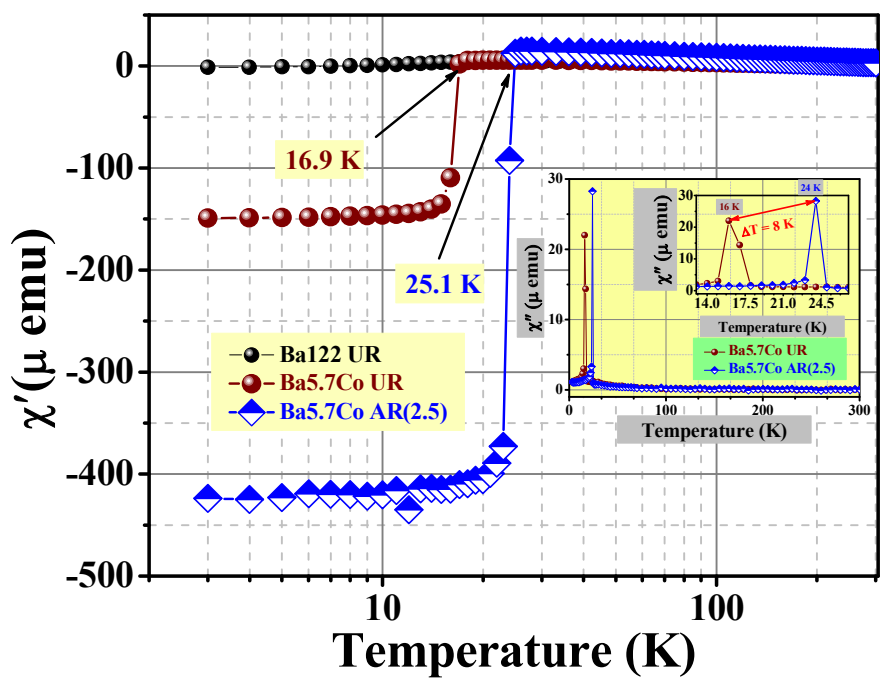

**Fig. 2:**

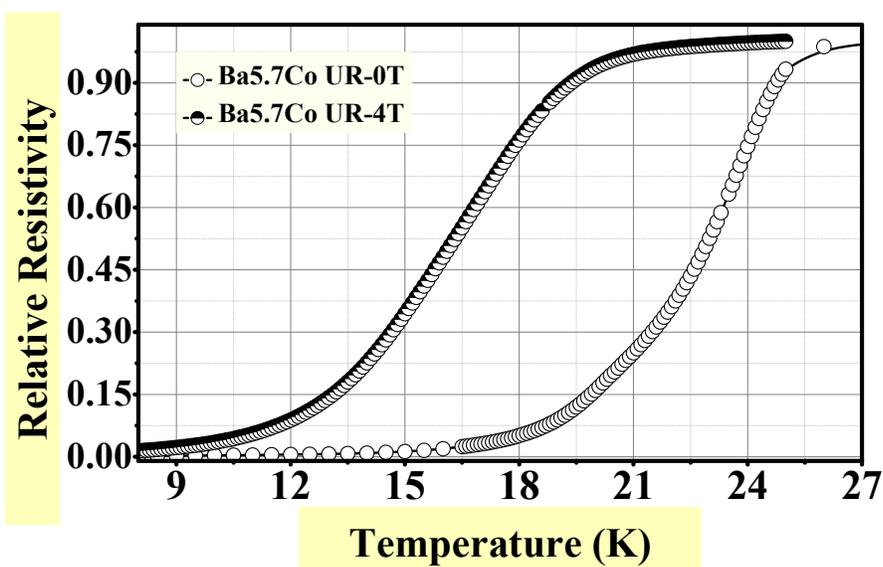

**Fig. 3:**

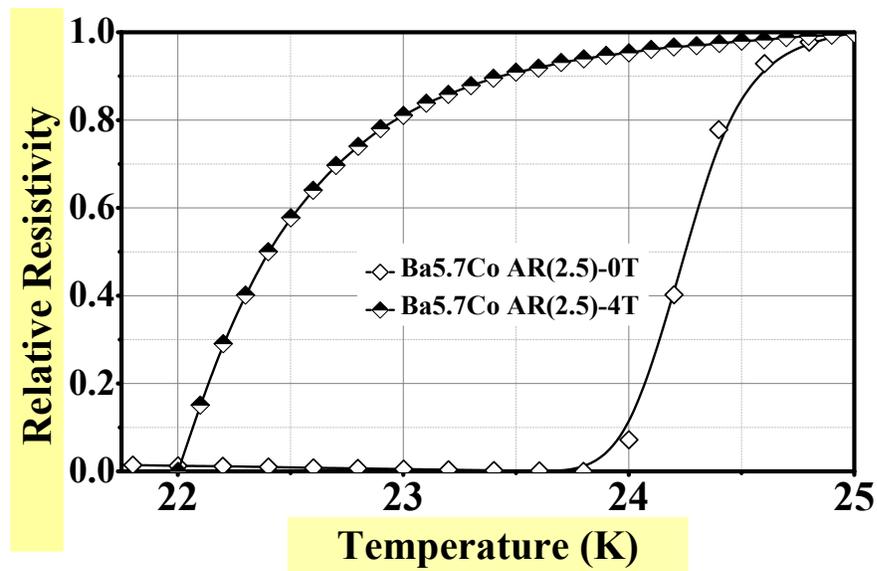

**Fig. 4:**

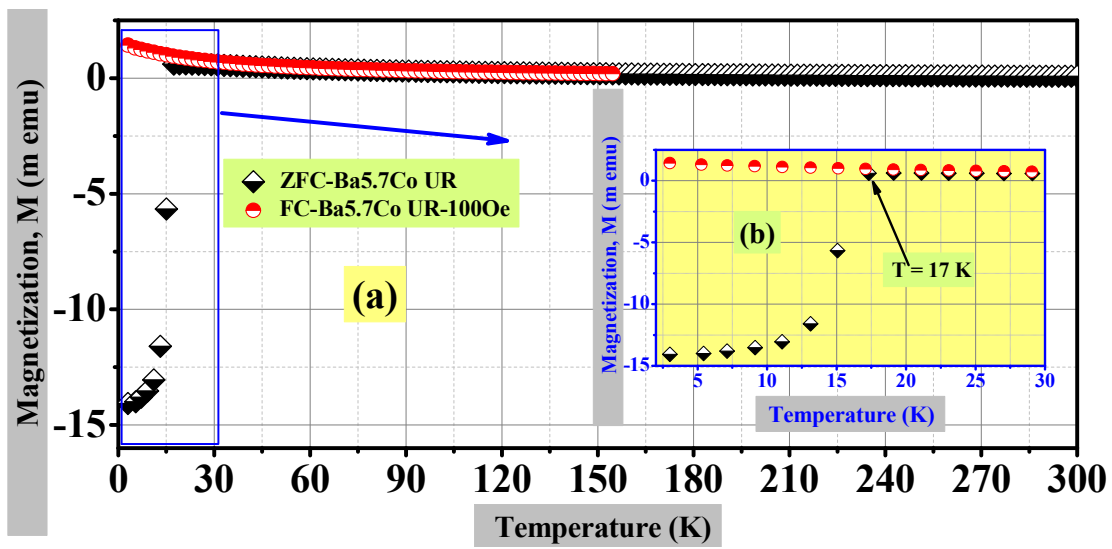

**Fig. 5**

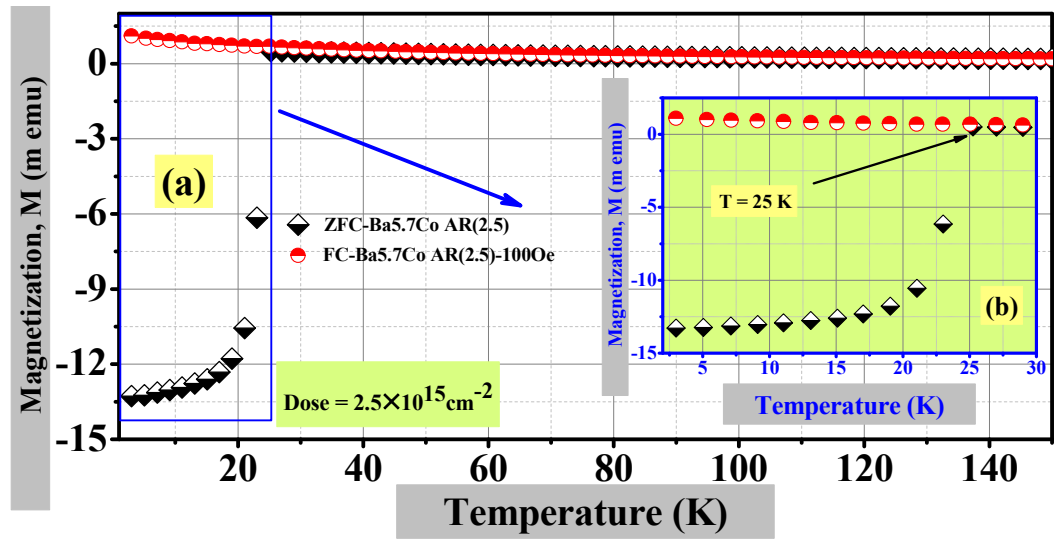

**Fig. 6**